\documentclass{PoS}

\newcommand{\be}{\begin{equation}}
\newcommand{\ee}{\end{equation}}

\newcommand{\bea}{\begin{eqnarray}}
\newcommand{\eea}{\end{eqnarray}}

\title{Fate of a recent conformal fixed point and $\beta$-function in the SU(3) BSM gauge theory with ten massless flavors}

\ShortTitle{Fate of an SU(3) 10-flavor fixed point}

\author{Zoltan Fodor\\
        University of Wuppertal, Department of Physics, Wuppertal D-42097, Germany\\
        Juelich Supercomputing Center, Forschungszentrum Juelich, Juelich D-52425, Germany\\
        Eotvos University, Pazmany Peter setany 1, 1117 Budapest, Hungary\\
        \email{fodor@bodri.elte.hu}}

\author{Kieran Holland\\
        University of the Pacific, 3601 Pacific Ave, Stockton CA 95211, USA\\
        \email{kholland@pacific.edu}}

\author{Julius Kuti\\
        University of California, San Diego, 9500 Gilman Drive, La Jolla, CA 92093, USA\\
        \email{jkuti@ucsd.edu}}

\author{\speaker{Daniel Nogradi}\\
        Eotvos University, Pazmany Peter setany 1, 1117 Budapest, Hungary\\
        \email{nogradi@bodri.elte.hu}}

\author{Chik Him Wong\\
        University of Wuppertal, Department of Physics, Wuppertal D-42097, Germany\\
        \email{cwong@uni-wuppertal.de}}

\abstract{SU(3) gauge theory with $N_f$ fermions in the fundamental representation serves as a theoretical testing ground for possible infrared conformal behavior, which could play a role in BSM composite Higgs models. We use lattice simulations to study the 10-flavor model, for which it has been claimed there is an infrared fixed point in the gauge coupling $\beta$-function. Our results suggest the opposite conclusion, namely we find no $\beta$-function fixed point in the explored range, with qualitative agreement with the 5-loop $\overline{MS}$ prediction. We comment on the inconsistency between our findings and other studies.}

\FullConference{The 36th Annual International Symposium on Lattice Field Theory - LATTICE2018\\
		22-28 July, 2018\\
		Michigan State University, East Lansing, Michigan, USA.}

\begin{document}

\section{Background}

\vspace{-2mm}
There is a long history of examining SU(3) gauge theory with varying number $N_f$ of massless flavors, to track the loss of spontaneous chiral symmetry breaking, the absence of a mass gap in the spectrum, and the emergence of conformality shown e.g.~by an infrared fixed point (IRFP) in the $\beta$-function. The study of $N_f = 8$ and $12$ in~\cite{Appelquist:2007hu} sparked renewed interest in the field. Theoretical developments and large-scale computing resources have since then brought significant improvement in accuracy. The general trend of the $\beta$ function decreasing in magnitude as $N_f$ increases is seen in many studies, with a corresponding drop in the composite scalar mass. Much of the focus has been on $N_f = 12$, with continuing discussion if the theory is IR conformal. Our results for $N_f = 12$ indicate that the $\beta$-function is small but non-zero, with no evident IRFP~\cite{Fodor:2016zil,Fodor:2017gtj}, contradicting other studies e.g.~\cite{Hasenfratz:2016dou,Hasenfratz:2018wpq}, the tension remains unresolved. If the 12-flavor model is not IR conformal, it is difficult to believe any smaller value of $N_f$ could be. Older studies of $N_f = 10$ using the Schr\"{o}dinger functional scheme claimed the model has an IRFP~\cite{Hayakawa:2010yn}, but with less statistical accuracy and at coarser lattice spacings than more recent work using a different scheme with the gradient flow and its associated renormalized coupling. One such study, with domain wall fermions, states the 10-flavor theory is IR conformal~\cite{Chiu:2017kza}, although systematic errors have since been reevaluated~\cite{Chiu:2018edw} with noticeable change in the results. We previously presented our simulation results that the $N_f = 10$ $\beta$-function shows no IRFP~\cite{Fodor:2017gtj} in the explored range of renormalized couplings, this work is an extension of that study.

\begin{figure}
\begin{center}
     \includegraphics[width=.48\textwidth]{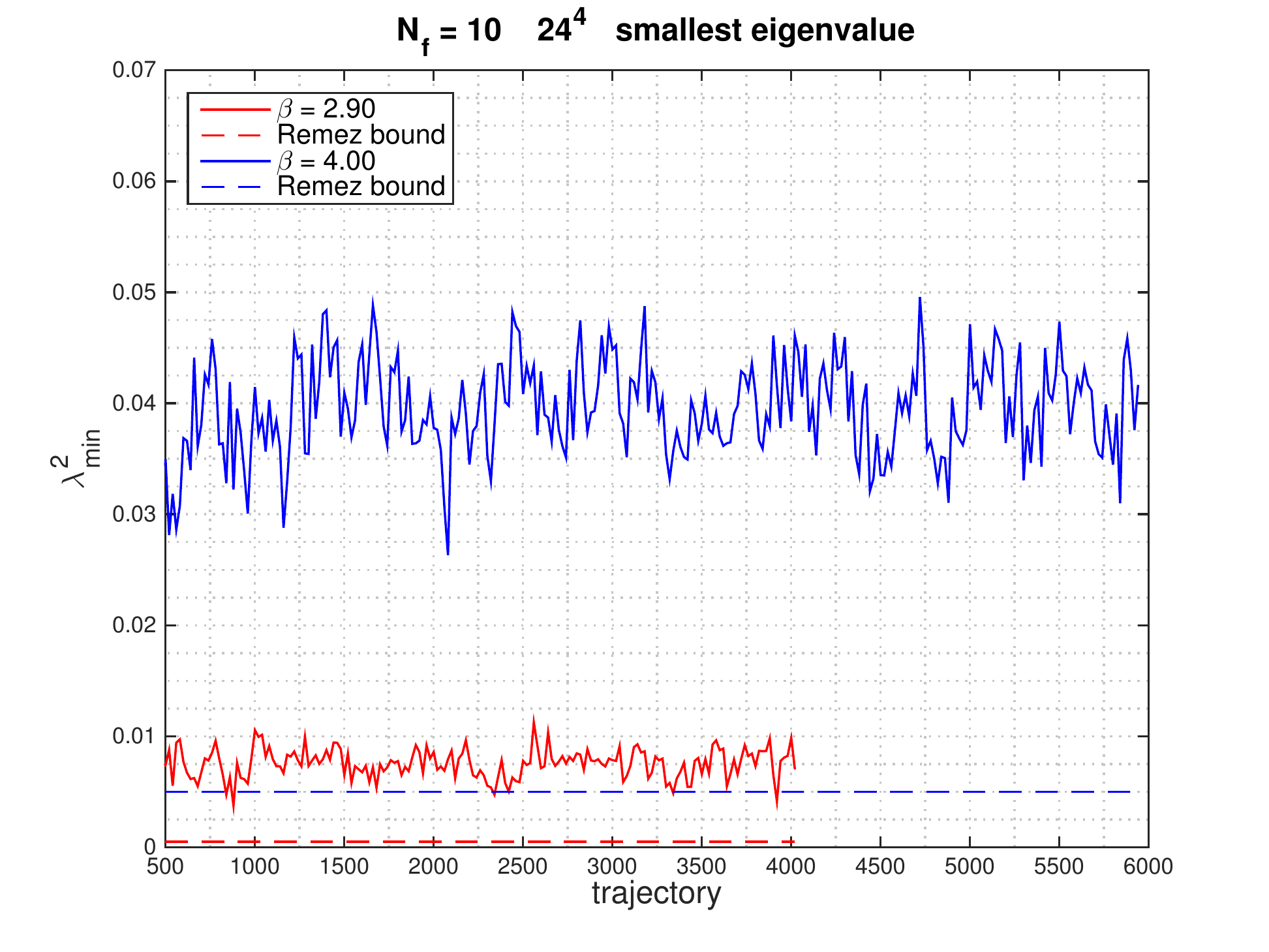}
     \includegraphics[width=.48\textwidth]{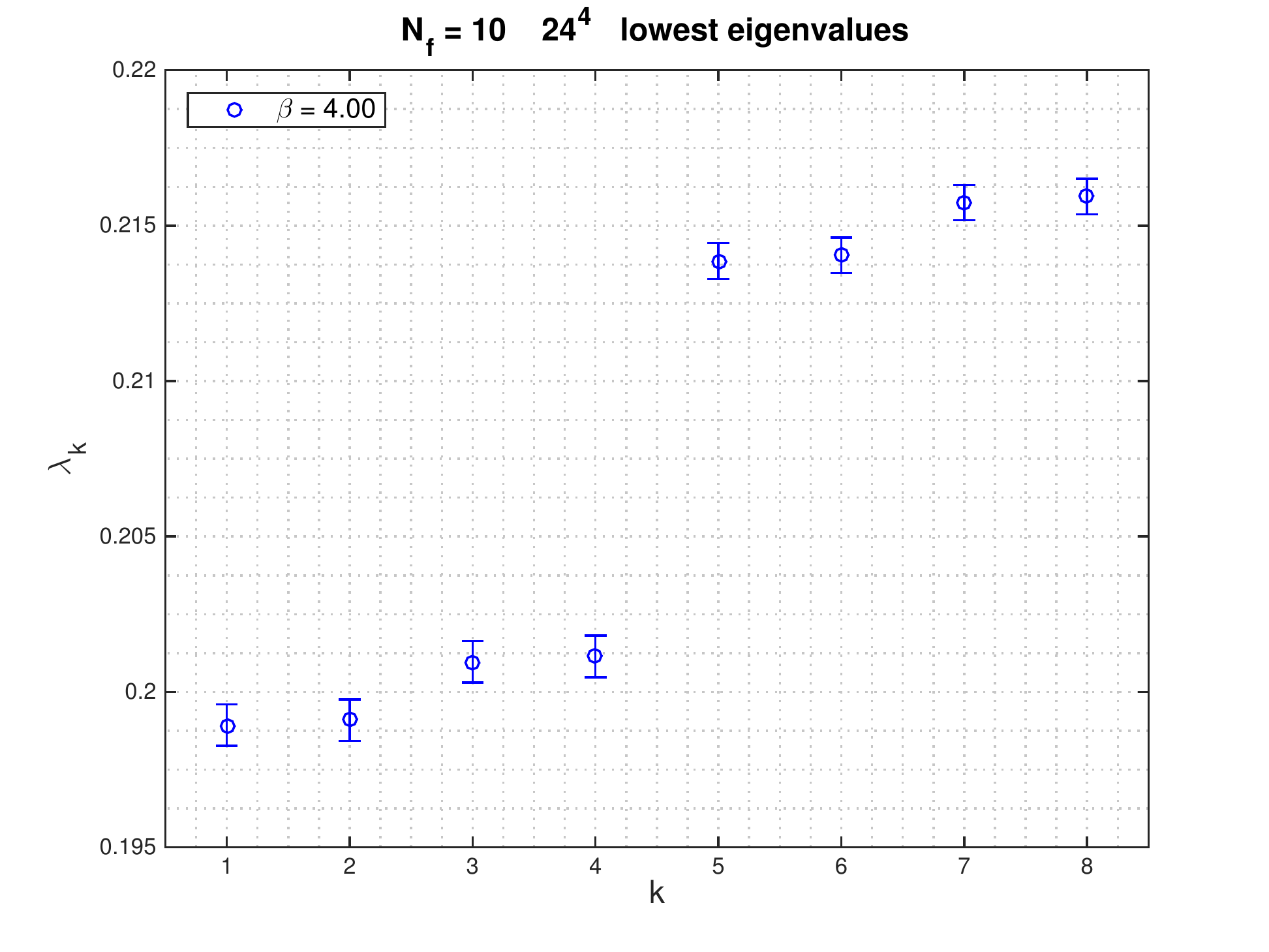}
     \caption{(left) The Monte Carlo history of the minimum eigenvalue of the Dirac operator, which is well above the chosen Remez bounds at strong and weak coupling. (right) The restoration of degenerate Dirac operator eigenvalue quartets towards the continuum limit, preceded by doublet formation.}
     \label{fig1}
\end{center}
\end{figure}

\vspace{-10mm}
\section{Lattice results}

\begin{figure}[hbt]
	\begin{center}
		\includegraphics[width=.48\textwidth]{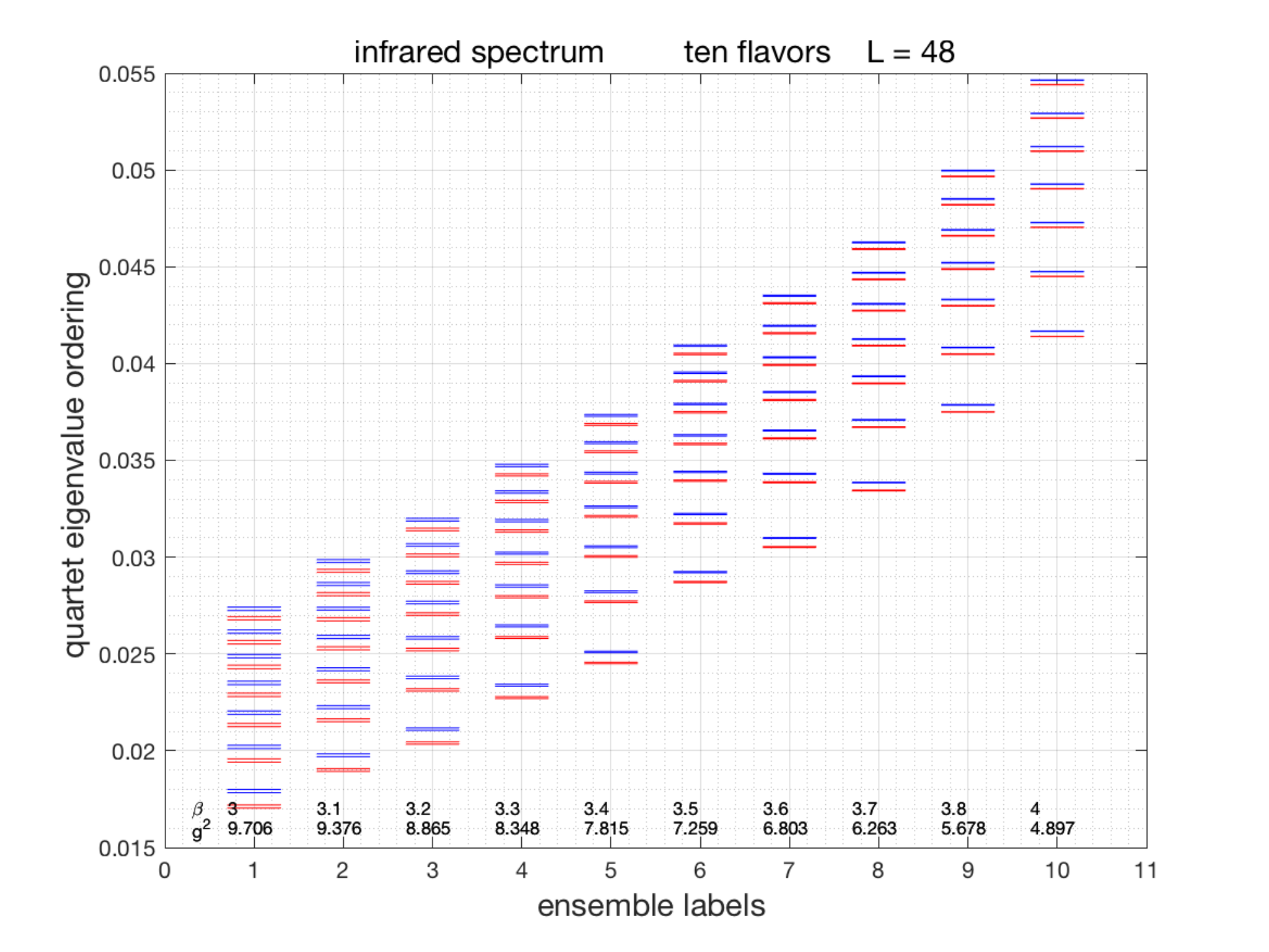}
		\includegraphics[width=.4\textwidth]{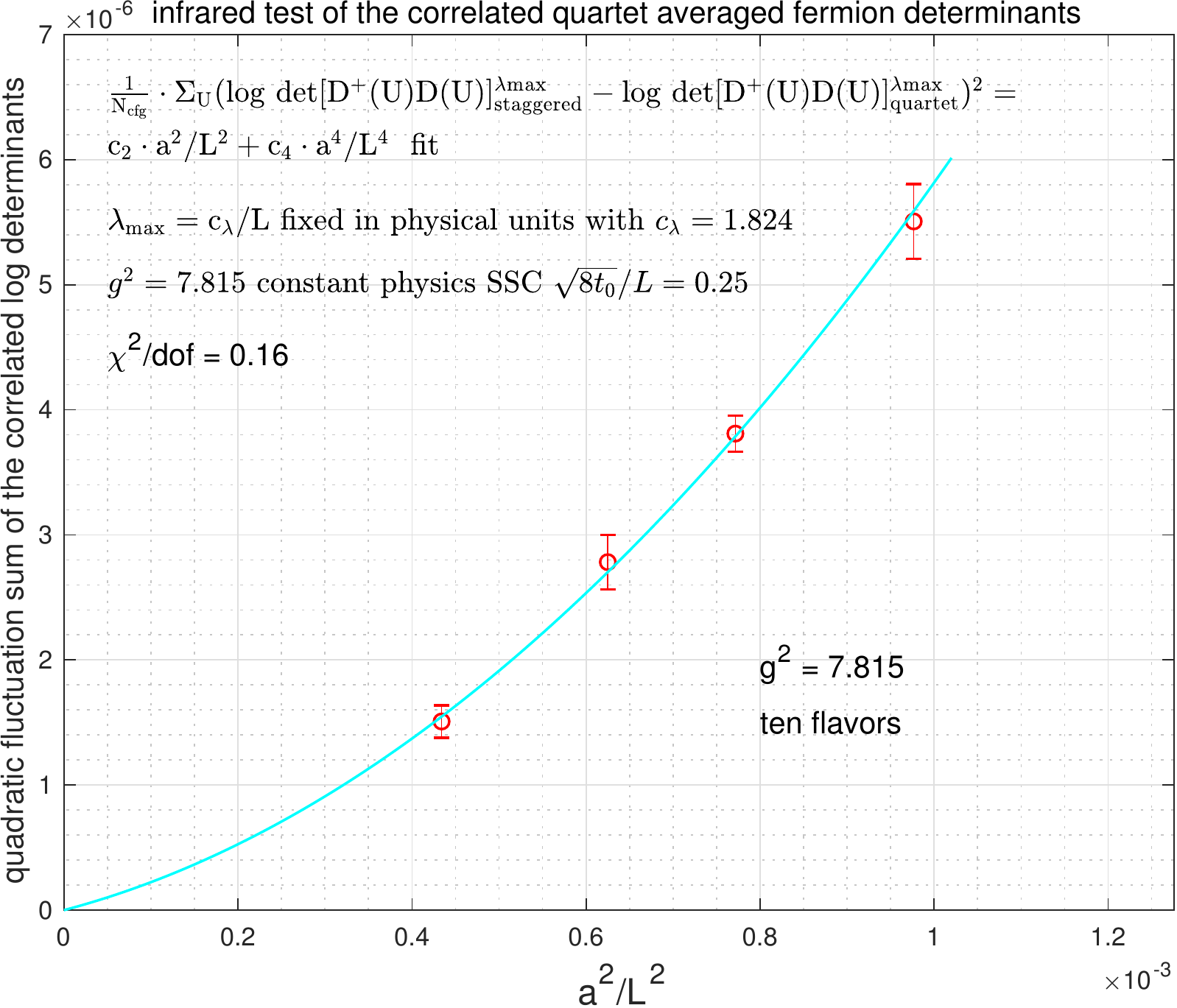}
		\caption{(left) Towers of the 28 lowest measured eigenvalues on $48^4$ volumes across the range of renormalized coupling. Red and blue lines denote eigenvalue pairs, which merge into quartets. (right) Taste breaking in the determinant of the Dirac operator at fixed renormalized coupling, where the eigenvalue cut is a fixed fraction of each lattice volume as defined in the text. It is clearly demonstrated that the measure of taste symmetry breaking vanishes in the continuum limit.}
		\label{fig2}
	\end{center}
	\vskip -0.2in
\end{figure}

\vspace{-2mm}
We simulate SU(3) gauge theory with $N_f = 10$ massless flavors using stout-smeared staggered fermions and the Symanzik-improved gauge action. We implement the flavor number with the RHMC algorithm, for which bounds on the Dirac operator eigenvalues must be set. We show in Fig.~\ref{fig1}  (left panel) typical Monte Carlo (MC) histories of the smallest Dirac eigenvalue at stronger and weaker coupling, the MC data are well above the chosen lower bounds. 
The Dirac eigenvalues do not form degenerate quartets due to taste-symmetry breaking. The emerging pattern is illustrated in Figs.~\ref{fig1} and ~\ref{fig2}.
The $N_f = 10~\beta$-function based on the staggered fermion formulation requires the square root of the fermion determinant in the functional integral to attain the correct flavor number. We have previously provided a comprehensive theoretical proof to validate this procedure in the presence of non-degenerate quartets~\cite{Fodor:2015zna}. Implicit in the proof is the restoration of quartet degeneracy in the continuum limit. Accordingly, at weaker gauge couplings pairs of doublets appear, which merge into degenerate quartets as $a \rightarrow 0$. 

We need to test how the staggered fermion determinant, built from split quartet eigenvalues at finite cutoff $a$, approaches the fermion determinant with exact quartet degeneracy in the continuum limit. Although we work in the staggered basis and the theoretical proof in~\cite{Fodor:2015zna} was built in the Dirac basis, once the quartet degeneracy is established in the staggered basis for the $a\rightarrow 0$  limit, the rooted theory is expected to become equivalent to the correct continuum fermion theory with the targeted flavor number. The staggered fermion determinant $\det~D_{\rm stagg}$ at finite $a$ can be viewed as the fourth power, $\det~D_{\rm stagg} = (\det~D_{\rm geom})^4$, of the determinant  $\det~D_{\rm geom}$  built from the geometric mean of split quartet eigenvalues with taste breaking. 
The reference determinant $\det~D_{\rm arith}$  is defined using the arithmetic mean of quartet eigenvalues. The ratio $\det~D_{\rm geom}/D_{\rm arith}$ is different from one unless the geometric and arithmetic means coincide. If the ratio is $1+{\cal O}(a^2)$ on gauge configurations, this validates the rooted staggered determinant in ensembles designed for the $\beta$-function at fixed physical volume in the continuum limit. 
The tests were performed on our $N_f =10$ $\beta$-function ensembles at five different physical volumes with corresponding renormalized gauge couplings $g^2 = 5.68, 6.26, 6.80, 7.26, 7.82$ defined with aspect parameter $c=0.25$ for the gradient flow. At each fixed physical volume and $g^2$ value, the infrared Dirac spectra were measured on four different lattice sizes with $L=32, 36, 40, 48$ up to a UV cutoff  $\lambda_{\rm max}$ set in physical units as $\lambda_{\rm max} = c_\lambda/L$ on the largest lattice and scaled up at smaller volumes using the well-known properties of the mode number distribution at fixed renormalized coupling
, similar to the anomalous mass dimension $\gamma$ determination in~\cite{Fodor:2018uih}.
The infrared part of the Dirac spectrum at $L=48$ is shown in Figure~\ref{fig2}.
The fluctuating  $\det~D_{\rm geom}/D_{\rm arith}$ ratios were measured on individual configurations with quadratic averages over the gauge ensembles. The cutoff dependence of the logarithmic ratios was measured and fitted in $a^2/L^2$ and $a^4/L^4$ orders as the lattice size $L$ varied over four values as shown in Figure~\ref{fig2} at the strongest coupling we targeted. The fits illustrate that taste breaking effects vanish in the continuum limit -- the correct Dirac determinant is recovered, giving the correct weight in the functional integral for MC simulations. This validates the use of rooting for staggered fermions and, in conjunction with other arguments, counters assertions that staggered fermions could be in the wrong universality class~\cite{Hasenfratz:2017mdh,Hasenfratz:2017qyr}. Although the tests were done on the infrared part of the fermion determinant, using our Chebyshev approximation for a high precision determination of the spectral function on all scales would extend the tests to the full determinant~\cite{Fodor:2016hke}.

\begin{figure}
\begin{center}
     \includegraphics[width=.48\textwidth]{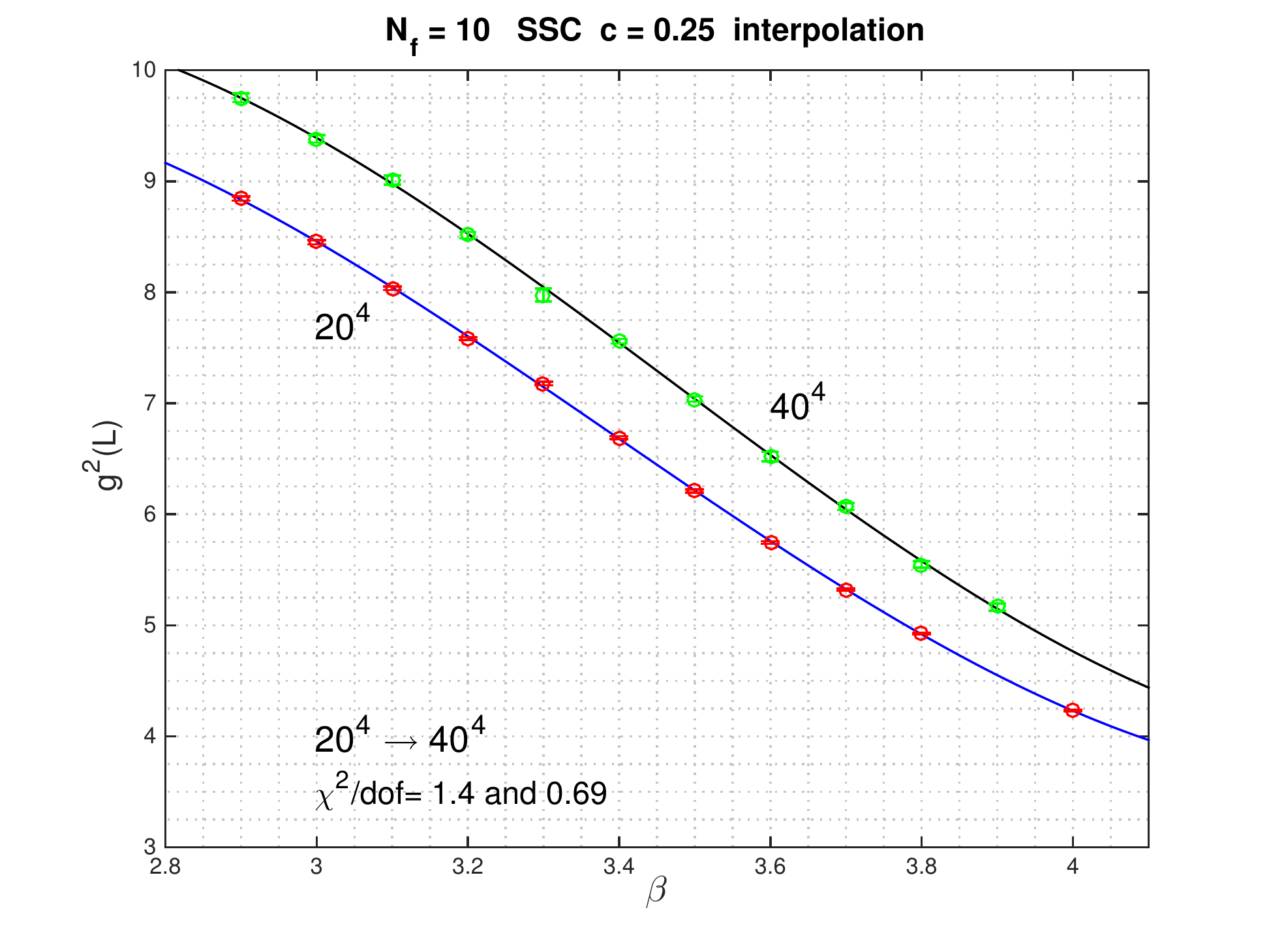}
     \includegraphics[width=.48\textwidth]{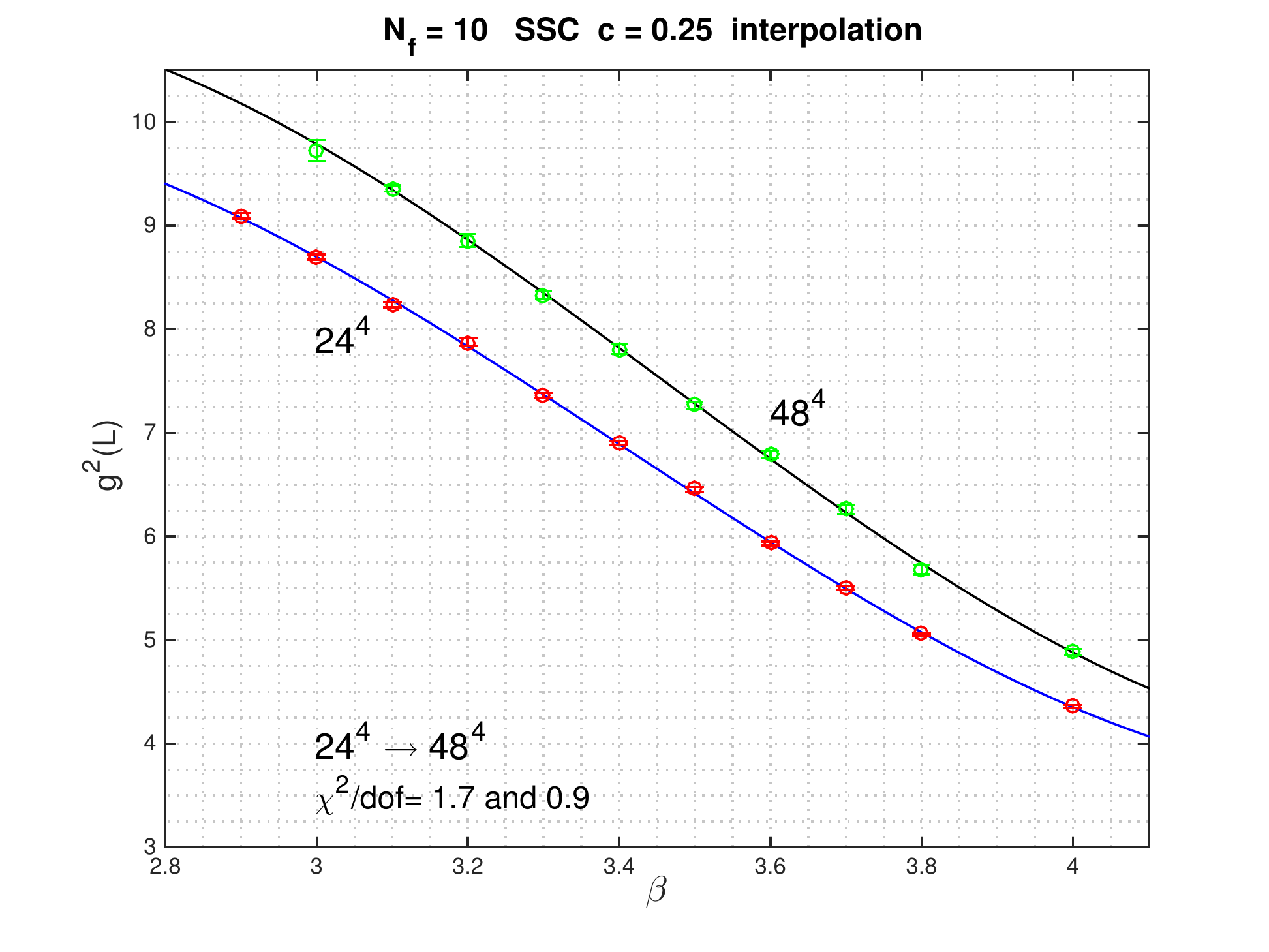}
     \caption{(left) Polynomial interpolation up to ${\cal O}(\beta^3)$ of the gradient flow renormalized coupling $g^2(L)$ in the bare coupling $\beta$ for volumes $20^4$ and $40^4$. Note the clear gap between the two volumes across all values of $\beta$. (right) The same interpolation on the largest pair of volumes $24^4 \rightarrow 48^4$.}
     \label{fig3}
\end{center}
\end{figure}

\phantom{.}
\vspace{-10mm}
We generate a range of gauge ensembles such that five lattice spacings are available for the discrete-step $\beta$-function $(g^2(sL) - g^2(L))/\log(s^2)$ with $s=2$, namely 
$L = 12/16/18/20/24 \rightarrow sL = 24/32/36/40/48$. This allows for a number of tests of lattice artifacts. The domain wall study~\cite{Chiu:2018edw} has $L = 8/10/12/16$ for the smaller volumes, the work~\cite{Hasenfratz:2017mdh} utilizes $L = 6/8/10/12$ (omitting the smallest for continuum extrapolations), so fewer and coarser lattice spacings in comparison. The discrete step is measured from paired lattices at the same bare coupling. To achieve a fixed renormalized coupling on all smaller volumes requires expensive tuning of the bare coupling. In its place we interpolate $g^2$ on each volume as a function of the bare coupling, shown in Fig.~\ref{fig3}, with a polynomial function. Displayed is the choice $c = \sqrt{8t}/L = 0.25$ for the finite-volume scheme. The split between $g^2(L)$ and $g^2(sL)$ is clearly visible on these largest volumes and hence smallest lattice spacings, a first hint that the continuum $\beta$-function is significantly far from  zero.

\begin{figure}
\begin{center}
     \includegraphics[width=.48\textwidth]{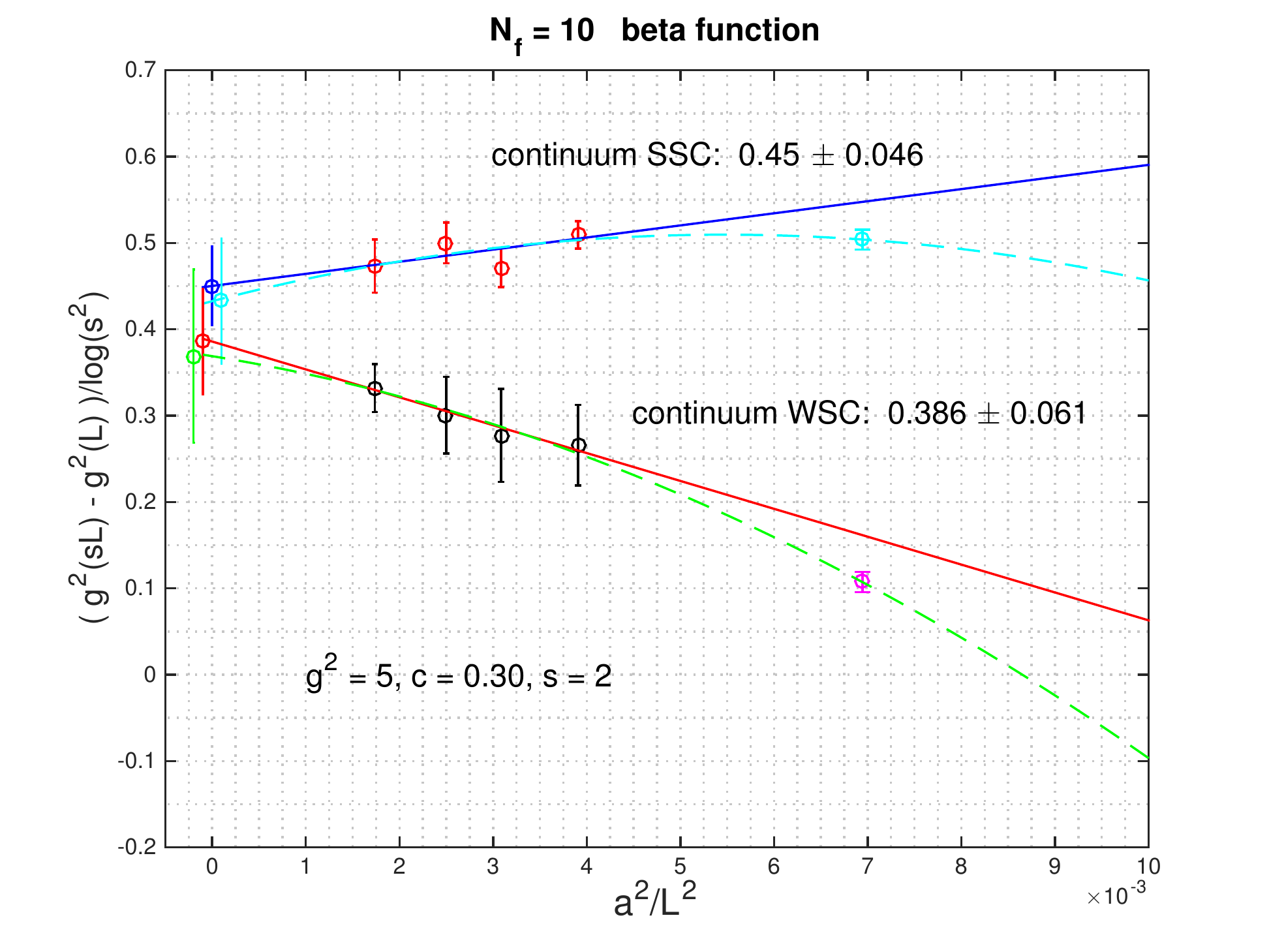}
     \includegraphics[width=.48\textwidth]{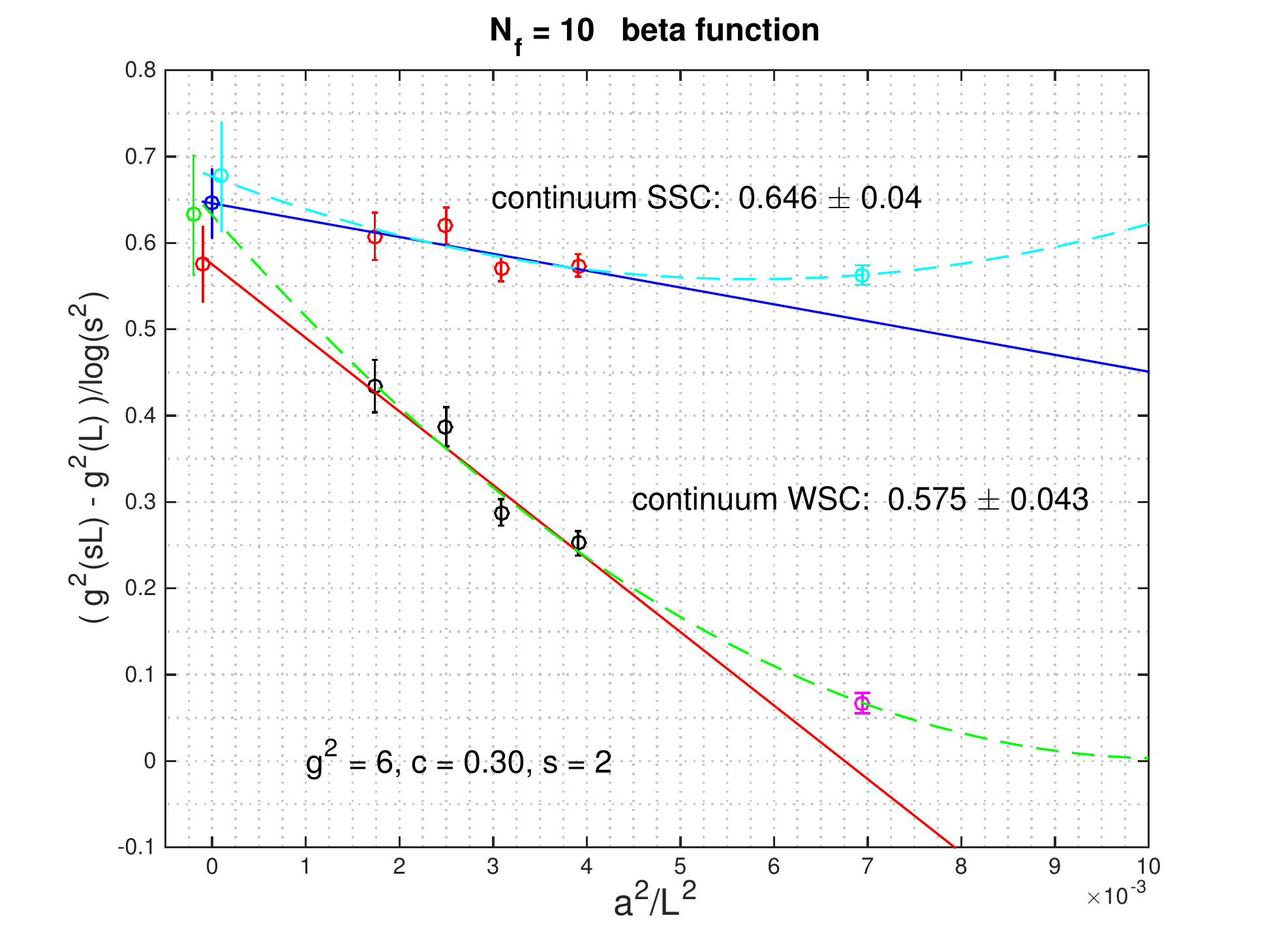}
     \caption{(left) Continuum extrapolation of the discrete $\beta$-function at fixed renormalized coupling $g^2 = 5$ with data at five lattice spacings: $12 \rightarrow 24, 16 \rightarrow 32, 18 \rightarrow 36, 20 \rightarrow 40$ and $24 \rightarrow 48$. The coarsest lattice spacing is included for quadratic in $a^2/L^2$ fits (dashed curves) and excluded for linear in $a^2/L^2$ ones (solid lines). Different discretizations of the gradient flow, namely Symanzik (S) or Wilson (W), yield consistent continuum results. The quoted continuum values are from the linear fits. (b) Similar continuum extrapolations at $g^2 = 6$. Note the increase in cutoff effects for the Wilson flow compared to $g^2 = 5$.}
     \label{fig4}
\end{center}
\end{figure}

We show in Fig.~\ref{fig4} examples of the continuum extrapolation at fixed $g^2$, where we choose the value $c= 0.30$ to allow direct comparison with the domain wall studies. The discretization variant SSC denotes Symanzik gauge action for the gradient flow and MC simulation, and Clover operator for the action density. Lattice artifacts with SSC are relatively mild, a quadratic extrapolation in $a^2/L^2$ of all 5 data or a linear fit excluding the coarsest point $L = 12$ give consistent continuum results, the quoted value is from the linear fit. The WSC discretization creates larger cutoff effects, but their independent fits agree well with the SSC ones and are a very useful crosscheck. Using the dashed quadratic fits as a guide, coarser lattice spacings such as $L = 8$ and 10 would be far removed from the linear scaling regime and not useful for continuum extrapolation.

\begin{figure}
\begin{center}
     \includegraphics[width=.48\textwidth]{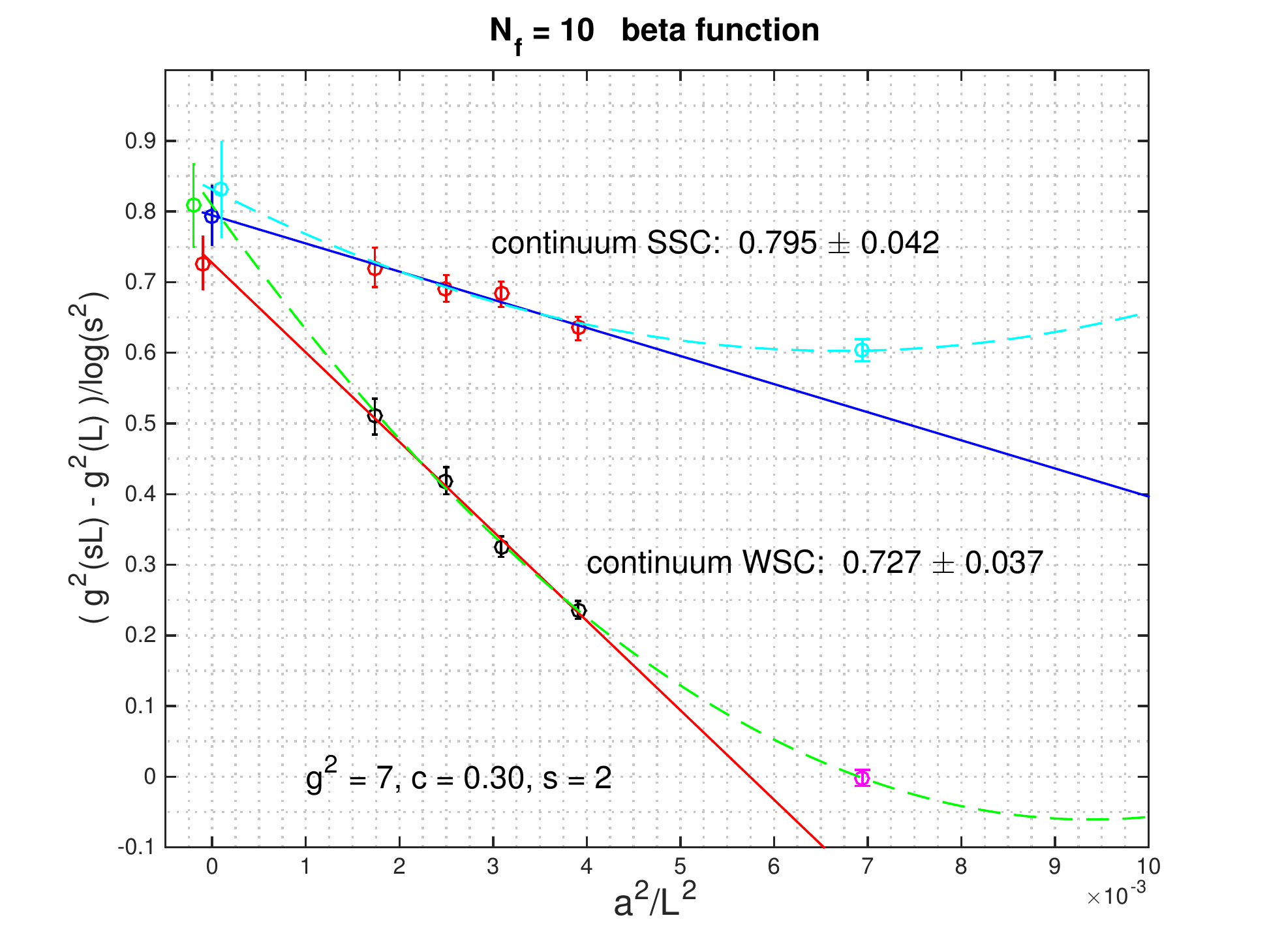}
     \includegraphics[width=.48\textwidth]{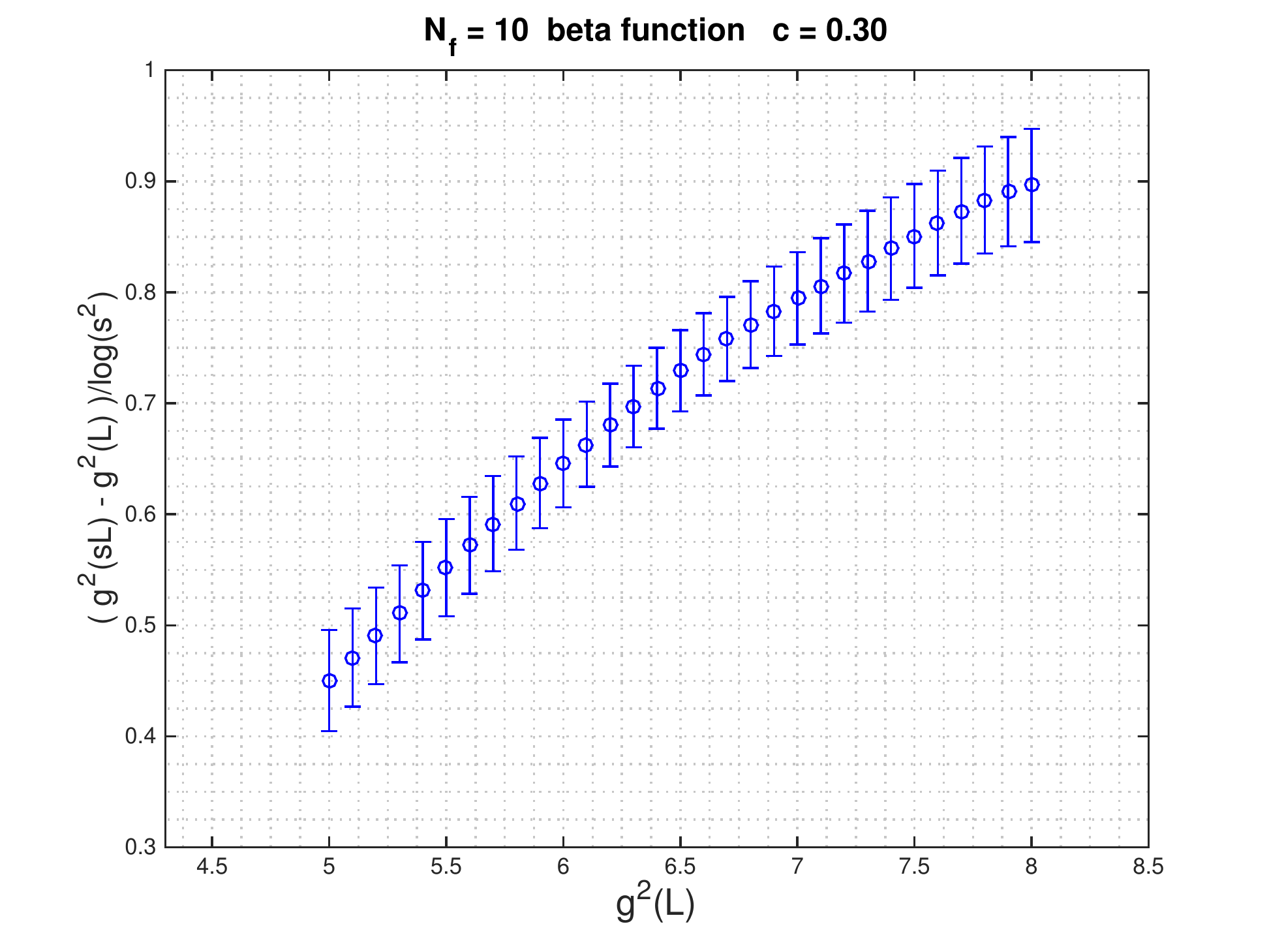}
     \caption{(left) Continuum extrapolation at renormalized coupling $g^2 = 7$. (right) The continuum discrete-step $\beta$-function for $c= 0.30$ and $s = 2$, given by linear continuum extrapolation. }
     \label{fig5}
\end{center}
\end{figure}

\begin{figure}
\begin{center}
     \includegraphics[width=.48\textwidth]{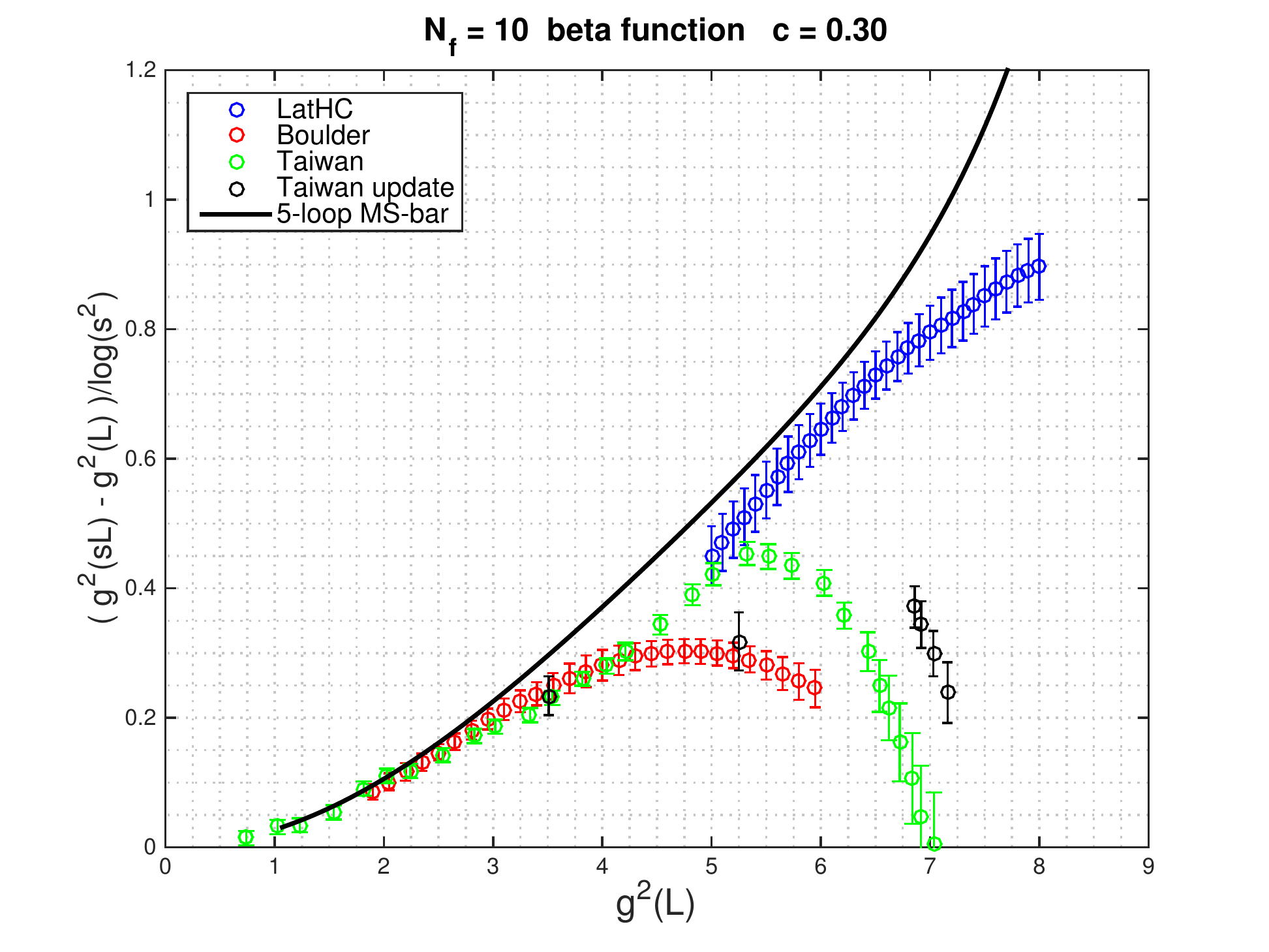}
     \includegraphics[width=.48\textwidth]{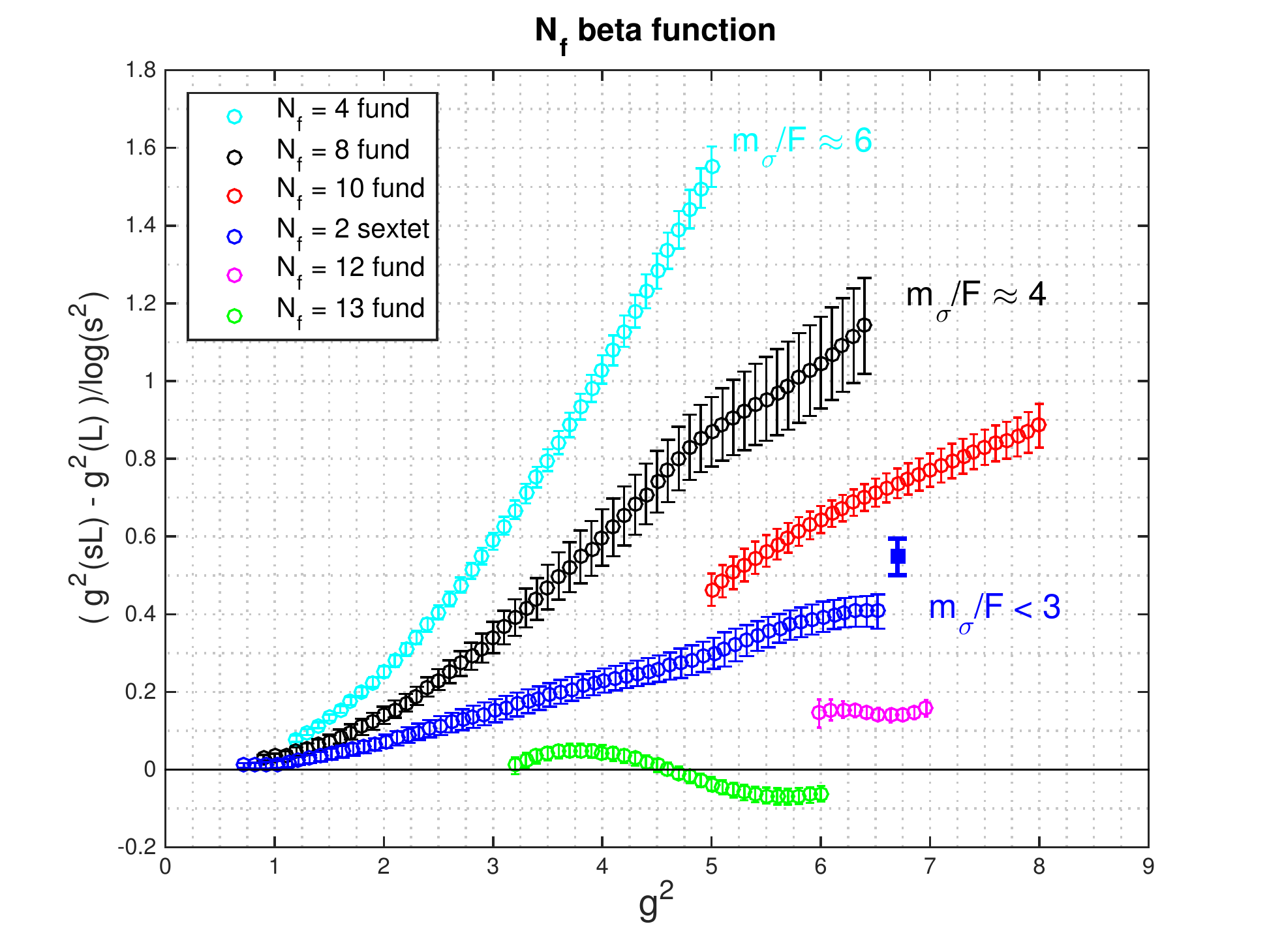}
     \caption{(left) Comparison of our result for $N_f = 10$ with \cite{Hasenfratz:2017mdh},~\cite{Chiu:2017kza} and \cite{Chiu:2018edw}, as well as the discrete-step $\beta$-function at 5-loop in the $\overline{MS}$ scheme. (right) A summary of the LatHC collaboration results for the $\beta$-function for a variety of SU(3) gauge theories, namely fundamental representation $N_f=4$~\cite{Fodor:2012td,Fodor:2014cpa}, 8~\cite{Fodor:2015baa}, 10~(this work), 12~\cite{Fodor:2016zil,Fodor:2017gtj}, 13~\cite{Fodor:2018uih} and sextet representation $N_f = 2$~\cite{Fodor:2015zna,Fodor:2017die}. We include where known estimates for the composite scalar mass $m_{\sigma}$ in units of $F$, the chiral limit of the pseudoscalar decay constant.}
     \label{fig6}
\end{center}
\end{figure}

\phantom{.} 
\vspace{-10mm}
There is a consistent trend on display in Figs.~\ref{fig4} and~\ref{fig5} -- the discrete $\beta$ function increases monotonically with $g^2$. Lattice artifacts with WSC discretization continue to grow, with SSC they are much smaller, nontheless the continuum results of the two variants are in good agreement. As shown in Fig.~\ref{fig6} our non-perturbative $\beta$-function results are quite close to the 5-loop prediction in the $\overline{MS}$ scheme~\cite{Baikov:2016tgj,Herzog:2017ohr} (the perturbative curve is also the discrete, not the infinitesimal, $\beta$-function). The results of~\cite{Chiu:2017kza} (green in the figure) and~\cite{Hasenfratz:2017mdh} (red), both using domain wall fermions, match closely one another up to $g^2 \sim 4.5$ and at weak coupling follow quite well the perturbative prediction. An update of these results since the conference~\cite{Chiu:2018edw} (black) shows a very significant shift in the $\beta$-function, the previous prediction of an IRFP at $g_\ast^2 \sim 7$ being changed to a possible IRFP at stronger coupling than the simulated range, with systematic errors in the interpolation of $g^2$ in the bare coupling cited as the cause. There remains a strong tension between those and our results. As the studies~\cite{Chiu:2017kza,Chiu:2018edw} use the single choice of Wilson flow and gauge action, and Clover operator (WWC) as the discretization, for which small tree-level cutoff effects without fermion feedback are not guaranteed for $N_f = 10$ dynamical fermions, there is no crosscheck of lattice artifacts with a different gradient flow. Our studies suggest that $L = 8$ and $10$, necessary in the analysis of the domain wall work, would be outside the scaling regime for a continuum extrapolation in our setup given the trends we see at finer lattice spacing.

The recent addition by our group~\cite{Fodor:2018uih} of an investigation of the $\beta$-function for $N_f = 13$, taken in conjunction with the results here, gives a very interesting overall summary of our collaboration's results, shown in the right panel of Fig.~\ref{fig6}. Starting from $N_f = 4$, with increasing flavor number the $\beta$-function has a steady decrease, while the composite scalar mass $m_\sigma$ drops, in units of $F$  the chiral limit of the pseudoscalar decay constant. This bolsters the hoped-for paradigm, that a Higgs impostor can emerge from the near-conformal particle spectrum. The fundamental representation allows more theoretical testing of the approach to conformality and possible dilaton-like behavior, given the broad range of $N_f$ values for which the theory is asymptotically free. In terms of few additional degrees of freedom and an exact match of Goldstone bosons to Electroweak gauge bosons the two-flavor sextet model remains interesting as an apparently near-conformal BSM candidate.

\vspace{-10mm}
\section{Summary}

\vspace{-2mm}
We present results that show the $N_f = 10$ theory has no IRFP in the range up to $g^2 \sim 8$, in contrast to the domain wall fermion simulations ~\cite{Chiu:2017kza,Chiu:2018edw}. Tests of the systematic effects of~\cite{Chiu:2017kza}, namely tuning versus interpolation, show that a putative IRFP would be shifted to larger coupling. One way to settle the discrepancy would be an independent determination of the $\beta$-function via simulations in the p-regime, using continuous gradient flow to measure the infinitesimal as opposed to discrete-step $\beta$-function, as described in \cite{Fodor:2017die} for the sextet model. Once this connection is made in the $\beta$-function between the weak and strong coupling phases, there is no freedom for an IRFP to emerge in the spontaneous chiral symmetry breaking regime on the strong coupling side. Of course continued study of the role of artifacts by comparing different discretizations will also play a role in resolving this question. 

\vspace{-3mm}
\acknowledgments
\vspace{-3mm}
We acknowledge support by the DOE under grant DE-SC0009919, by the NSF under grant 1620845, by NKFIH grant KKP-126769, and by the Deutsche Forschungsgemeinschaft grant SFB-TR 55. KH thanks the AEC at the University of Bern for their support. Computational resources were provided by the DOE INCITE program on the ALCF BG/Q platform, by USQCD at Fermilab, by the University of Wuppertal, and by the Juelich Supercomputing Center on Juqueen. We thank Szabolcs Borsanyi, Sandor Katz and Kalman Szabo for code development.

\end{document}